\documentstyle[preprint,aps]{revtex} 
\begin{document}

\draft

\title{A square root of the harmonic oscillator}

\author{R Delbourgo\thanks{In memory of Lorella M Jones who
sadly passed away in February 1995 and to whom the original idea
in this paper is due.}}

\address{Physics Department, University of Tasmania,\\
GPO Box 252C Hobart, Australia 7001}
\date{\today}

\maketitle

\begin{abstract}
Allowing for the inclusion of the parity operator, it is
possible to construct a model of an oscillator whose Hamiltonian
admits an exact square root which is rather different from the
conventional approach based on creation and annihilation
operators. We outline such a model, the method of solution
and some generalisations.

\end{abstract}


\narrowtext
\section{INTRODUCTION}

The simplest way of solving the harmonic oscillator uses
the fruitful concept of annihilation $A$ and creation $A^\dagger$
operators. $A$ and $A^\dagger$ are of course non-hermitian and
therefore differ from one another just like a complex number and
its conjugate; nonetheless, in a loose way, they can be construed
as providing a square root of the harmonic Hamiltonian,
since the number operator is given by $N=A^\dagger A$. The same
idea applies to supersymmetric models where the Hamiltonian
is expressed as a quadratic function of hermitian fermionic
operators $F_i$, {\em viz.} $\sum_{ij} c^{[ij]}F_iF_j.$ In this paper
we wish to exhibit a model for which there exists an exact
square root of the Hamiltonian that is not far removed from
the one for an oscillator. In that sense it represents a much
closer analogy with the Dirac equation, which is a square root
of the Klein-Gordon equation at the price of doubling the
number of components (namely introducing spin). Indeed,
we shall find something similar in the system described below.
To dispose of unwanted constants and thereby simplify the
algebra, let us set the mass and circular frequency $\omega$ of the
oscillator equal to 1, as well as $\hbar$.

It is worth mentioning that a number of papers\cite{ms,mz,b,c} have
appeared concerning the Dirac oscillator. These correspond to making
a minimal substitution $\vec{p}\rightarrow\vec{p}-i\beta\vec{r}$ in
the free Dirac equation and they can be regarded as placing the
spinor particle in a linearly increasing potential. We should stress
that our approach is quite different from those papers since it
even applies to a bosonic system. The crucial respect where we
differ is our reliance upon the parity operator $Q$.

The basic point is to notice that whereas $A = P + iX$
is non-hermitian, the combination $D = P + iQX$ is real\cite{f1}.
Here $P$ stands for the momentum operator, $X$ is the position
operator and we are working in one spatial dimension for the
moment. Using the fundamental
commutators,
$$ [X,P] = i, \qquad \{X,Q\} = \{P,Q\} = 0,\qquad Q^2=1,$$
and squaring $D$, one gets
\begin{equation}
 H\equiv D^2=P^2+X^2-Q =2H_{\rm osc}-Q.
\end{equation}
Thus apart from a parity term $Q$ \cite{f2}, which commutes with
$H_{\rm osc}$, we get an exact square root of the oscillator.
It is an unexpected and striking result that we do not require
non-hermitian operators for `square-rooting' model (1).

In the next three sections, we shall study the eigenvalue equation
for $D$, extend the idea to higher dimensions and generalise the
analysis to the Dirac equation. A full set of physical
solutions will be determined which do not depend on an external
vector; we shall also investigate zero eigenvalues that are
characterised by some external (null) wavenumber.

\section{Eigenfunctions of $D$}

Because the two parts of $H$ behave differently under parity, we may
anticipate that a doubling of components is needed and that is
precisely what the following representation of the various operators
in $H$ gives,
\begin{equation}
 P \rightarrow -i\sigma_1\,\partial/\partial x, \qquad
 X \rightarrow \sigma_1 x, \qquad Q\rightarrow\sigma_2.
\end{equation}
Thus $D$ is represented by the matrix,\cite{f3}
$$D = \rightarrow -i\sigma_1\,\partial/\partial x + \sigma_3 x
 = \left( \begin{array}{cc} x & -i\partial/\partial x \\
                          -i\partial/\partial x & -x
          \end{array} \right). $$
Letting $(f(x),g(x))$ denote the eigenfunctions of $D$, it pays
to form the combinations $f_\pm \equiv f\pm ig$, corresponding
to projections $(1\mp Q)$, so as to obtain the two coupled first
order equations,
\begin{equation}
 (\pm x + d/dx)f_\mp = \pm\lambda f_\pm,
\end{equation}
where $\lambda$ connotes an eigenvalue of $D$.  In turn these can
be converted into uncoupled second order equations,
\begin{equation}
 f_\pm\,^{\prime\prime} - (x^2 -\lambda^2 \pm 1)f_\pm = 0.
\end{equation}

The standard method of solution, which is to put $f_\pm = F_\pm
\exp(-x^2/2)$ and work out the (terminating) series expansion of
$F_\pm$, shows that the excited states are
$$f_+\propto H_N\exp(-x^2/2),\quad f_-\propto\pm
    H_{N+1}\exp(-x^2/2);$$
with $\lambda=\pm\sqrt{2(N+1)}, N=0,1,\ldots$. intrinsic parity
$Q=(-1)^{N+1}$; in contrast, the ground state is
$$f = -ig \propto \exp(-x^2/2); \qquad \lambda = 0,$$
with intrinsic parity $Q=1$. Here $H_N(x)$ signify the usual Hermite
functions. Altogether then, with proper normalization, we write
the (doubled) excited states in the form
\begin{equation}
 \psi_{N}(x) = \frac{-i\exp(-x^2/2)}{\sqrt{2^{N+1}(N-1)!\sqrt{\pi}}}
             \left(\begin{array}{c}
                    i(H_{N-1}\pm H_N/\sqrt{2N}) \\
                     (H_{N-1}\mp H_N/\sqrt{2N})
                   \end{array} \right),
\end{equation}
with $\lambda = \pm\sqrt{2N}$; $N=1,2,\ldots$. The ground state
($\lambda=0$) is
\begin{equation}
 \psi_0(x) = \frac{\exp(-x^2/2)}{\sqrt{2\sqrt{\pi}}}
          \left( \begin{array}{c} 1 \\ i \end{array}\right).
\end{equation}
As a matter of fact one may regard the ground state as the
continuation of (5) to $N=0$, remembering that $(-1)!=\infty$.

One can then go on to determine expectation values, uncertainties
and such like, using well-known properties of Hermite functions.
For instance, the ground state quite obviously yields,
$$\langle X \rangle = \langle P \rangle = 0,\quad
  \Delta X. \Delta P = 1/2, $$
whereas the excited states ($N=1,2,\ldots$) have
$$\langle X \rangle = 0, \quad \langle P \rangle = \pm\sqrt{N/2},
  \quad \Delta X. \Delta P = N/\sqrt{2}, {\rm etc.} $$
However instead of elaborating upon this 1-dimensional model, it
is much more interesting to examine its higher-dimensional
generalizations.

\section{Higher Dimensions}

With several independent coordinates ($j=1,\ldots,D$) we can
contemplate at least two distinct choices for the square roots
$D_j$, viz. $P_j + iQ_jX_j$ {\em or} $P_j+iQX_j$, where $Q_j$
reflects $x_j$ alone, whereas $Q=\prod_jQ_j$ is the full
inversion operator. The problem with the first choice is quite
serious: the components $D_j$ do not transform as a true vector
under rotation, in contrast to the second choice. However in the
latter case one should observe that
\begin{eqnarray}
 [D_j,D_k] &=& [P_j+iQX_j,P_k+iQX_k]\nonumber \\
           &=& 2iQ(X_jP_k-X_kP_j) = 2iQL_{jk}.
\end{eqnarray}
Nevertheless it remains true that $\sum D^2$ is effectively the
$D$-dimensional oscillator, barring a parity term:
$$D_jD_j = P_jP_j + X_jX_j - DQ = 2H_{\rm osc} - DQ. $$
We will therefore stick with the second choice.

In order to determine the eigenfunctions of this Hamiltonian, we
cannot take simultaneous eigenfunctions of the $D_j$ separately
since the components will not commute. Still, we may construct
joint eigenfunctions of $D_jD_j, Q$ and $L^2$, where $L_{jk}$
are the $D(D-1)/2$ components of the angular momentum tensor.
Now in $D$-dimensions, proceeding to appropriate spherical
coordinates,
$$\psi(x) = R(r)Y(\hat{x}), \qquad
  L_{jk}L_{jk}.Y(\hat{x}) = 2\ell(\ell + D -2).Y(\hat{x}), $$
the radial equation for $D^2 \psi = \lambda^2\psi$ can be reduced
to ${\cal O}_rR(r) = 0$, where
\begin{equation}
{\cal O}_r\equiv\frac{d^2}{dr^2} + \frac{D-1}{r}\frac{d}{dr} -
      \frac{\ell(\ell+d-2)}{r^2} - r^2 + \lambda^2 + DQ.
\end{equation}
As ever we look for a solution, $R(r)={\cal H}(r)\exp(-r^2/2)$
where $\cal H$ is a terminating power series in $r$ and obeys
${\cal D}_r{\cal H}(r) = 0$, with
$${\cal D}_r=\frac{d^2}{dr^2} + (\frac{D-1}{r} - 2r)
           \frac{d}{dr}+\lambda^2+DQ-D-
               \frac{\ell(\ell+D-2)}{r^2} $$
It is an elementary exercise in differential equations to
prove that ${\cal H}(r) = \sum_j c_jr^{j+\ell}$ with $c_0\neq0$
where the coefficients satisfy the recurrence relation,
$$(j+2)(j+2\ell+D)c_{j+2} = (2j+2\ell+D-DQ-\lambda^2)c_j. $$
Clearly we must set the odd terms, starting with $c_1$, equal to 0
and require that
$$\lambda^2 = 4n_r + 2\ell + D -DQ; \qquad (Q=\pm 1), $$
where the radial quantum number $n_r = 0,1,2,\ldots$ has the
same range of integer values as $\ell$. In short we have found the
complete solution to this problem as a `radial Hermite' function
multiplying a spherical harmonic and radial gaussian, whose
coefficients are fixed by
$$\frac{c_{j+2}}{c_j} = \frac{2(j-2n_r)}{(j+2)(j+2\ell+d)}. $$
Finally, the eigenvalues $\lambda^2$ of $D^2$ are given by
\begin{equation}
 \lambda^2 \equiv 2N = 4(2n_r+\ell) +(D\mp D); \qquad Q=\pm 1.
\end{equation}

It is tempting to suppose that the above discussion applies to
the relativistic Klein-Gordon operator in the particular case
where $D\rightarrow 4$ and we continue in the time coordinate,
$x_4 \rightarrow it, p_4 \rightarrow iE$. Although it is true
that the solutions
\begin{equation}
 \psi_N(x) = {\cal H}(\rho).Y_\ell(\hat{\rho}) \exp(-\rho^2/2);
 \qquad \rho^2 \equiv -x^2,
\end{equation}
with
$$\lambda^2 = 4n_r+2\ell+\left(\begin{array}{c}0\\8\end{array}
                    \right), \qquad {\rm for}\quad Q=\pm 1,$$
are certainly normalizable over 3-space at $t=0$, they have
unacceptable behaviour in the distant past or future and
share some of the time-dependence problems of Bethe-Salpeter
wavefunctions that are not evaluated in the instantaneous
approximation. We therefore somewhat hesitant to claim\cite{kn} that
such a covariant Hamiltonian ($=D_\mu D^\mu$) with its
Gaussian factor $\exp(x^2/2)$ makes good physical sense.

\section{Another Dirac Oscillator}
We are now going to extend the discussion covariantly to the Dirac
equation, but differently from refs\cite{ms,mz,b,c}. Continuing to
define $Q$ as the total inversion ($x\rightarrow -x$) operator, the
Dirac oscillator equation
\begin{equation}
 \gamma.(P + iQX)\psi = i\lambda\psi,
\end{equation}
can be represented in the doubled 4-component form
$\gamma.(\partial\sigma_1+ix\sigma_3)\psi = \lambda\psi$, or
\begin{equation}
 \left(\begin{array}{cc}
       -\gamma.x & i\gamma.\partial \\
       i\gamma.\partial & \gamma.x \end{array}\right)
 \left(\begin{array}{c}f \\ g \end{array}\right) =
 i\lambda\left(\begin{array}{c}f \\ g \end{array}\right).
\end{equation}
It helps to form the combinations $f_\pm \equiv f\pm ig$ as before,
so that the equations simplify to the coupled pair
\begin{equation}
 \gamma.(\partial \pm x)f_\mp = \mp i\lambda f_\pm.
\end{equation}
``Squaring'', we arrive at the uncoupled pair of differential
equations,
\begin{equation}
 (\partial^2 - x^2 \pm 4 \pm \sigma^{\mu\nu}L_{\mu\nu})f_\mp =
 \lambda^2 f_\mp,
\end{equation}
since $\not x\not\!\!\partial-\not\partial\!\!\not x
  =\gamma^\mu\gamma^\nu (x_\mu\partial_\nu - \partial_\mu x_\nu)
  =-4-\sigma^{\mu\nu}L_{\mu\nu}$
with $L_{\mu\nu} \equiv i(x_\mu\partial_\nu - x_\nu\partial_\mu)$.

Now products of spinors and 4-dimensional spherical harmonics
decompose relative to the Lorentz group as the sum of two
parity conserving irreducible representations $A$ and $B$
$$[(\frac{1}{2},0)\oplus(0,\frac{1}{2})]  \otimes
   (\frac{\ell}{2},\frac{\ell}{2}) = [A] \oplus [B], $$
where
$$A \equiv (\frac{\ell+1}{2},\frac{\ell}{2}) \oplus
           (\frac{\ell}{2},\frac{\ell+1}{2}), $$
$$B \equiv (\frac{\ell-1}{2},\frac{\ell}{2}) \oplus
           (\frac{\ell}{2},\frac{\ell-1}{2}). $$
The square of the total 4-dimensional angular momentum $J =
L + \sigma/2$ acts thus:
\begin{equation}
 J^{\mu\nu}J_{\mu\nu}.[A] = (2\ell^2 + 6\ell + 3).[A],
\end{equation}
whereupon it follows that
$$ \sigma^{\mu\nu}L_{\mu\nu}.[A] = 2\ell.[A]; $$
similarly for the wavefunction $[B]$, upon effecting the substitution
$\ell\rightarrow\ell - 1$.
Hence on one of these irreducible wavefunctions (labelled by $A$ say),
$f_{A+}(x) = F_{A+}(\rho).Y_A(\hat{\rho})$ our oscillator equation
simplifies to
\begin{equation}
 (\partial^2 - x^2 -4 - 2\ell - \lambda^2)f_{A+}(x) = 0.
\end{equation}
As usual, put $F_{A+}(\rho)={\cal H}(\rho)\exp(-\rho^2/2);\quad \rho^2
= - x^2$, to get a radial equation ${\cal O}_\rho{\cal H}(\rho)=0$,
having
$$ {\cal O}_\rho=\frac{d^2}{d\rho^2}+\left(\frac{3}{\rho}-2\rho\right)
                 \frac{d}{d\rho} + \left(\lambda^2+2\ell-
                       \frac{\ell(\ell+2)}{\rho^2}\right).$$
The (polynomial in $\rho$) eigenfunctions are readily found,
$${\cal H}_{A+}(\rho) = \sum_{j=0}^{n_\rho} c_j\rho^{j+\ell}, $$
with $\lambda^2=4n_\rho;\quad n_\rho=0,1,\ldots$ and
$$\frac{c_{j+2}}{c_j} = \frac{2j-4n_\rho}{(j+2)(j+2\ell+4)},\quad
  c_0\neq 0;\quad j{\rm ~even}.$$

One may carry out a similar analysis of the $f_{A-}$ equations; this
time one ends up with the eigenvalues $\lambda^2 = 4n_\rho+2\ell+8$.
Of course we get exactly the same answers for the $[B]$ eigenfunctions,
apart from the change $\ell \rightarrow \ell -1$. This then completes
the analysis when there is no dependence of $f,g$ on an external wave
vector.

Let us now have a brief look at the zero eigenvalue problem allowing
for some external wavenumber---which is therefore analogous to seeking
plane wave solutions of the ordinary free Dirac equation. Recall from
(13) that if $\lambda=0$, the first order equations decouple.
In particular the equation for $f_-(x)$ has asymptotic behaviour
$\exp(-x^2)/2$ which is certainly unacceptable in the space-like
direction; that is we are obliged to set $f_- = 0$. On the other hand,
putting $f_+(x)=F_+(x)\exp(x^2/2)$, one simply remains with
$\gamma.\partial F_+ = 0$, corresponding to a free massless equation.
Hence we arrive at the solution ($\lambda=0$),
$$f_+(x) = [u + \not k v \exp(-ik.x)\delta(k^2)]\exp(x^2/2). $$
In short, the ground state is fixed by a null wave number and may
be written
$$\left(\begin{array}{c} u + \not k v \exp(-ik.x)\delta(k^2) \\
      -i[u + \not k v \exp(-ik.x)\delta(k^2)]\end{array} \right)
   \exp(x^2/2),$$
where $u$ and $v$ are constant 4-spinors. We have not succeeded in
determining solutions for $\lambda\neq 0$ that are specified in terms
of a non-null $k$, as the equations are much more complicated in that
case.

Finally, observe that what progress one has achieved depends
critically on the fact that there is a double degeneracy
when the eigenvalue $\lambda$ is nonzero, but a single degeneracy
for the ground state $\lambda = 0$; this happy accident has arisen
because the Hamiltonian is exactly that of the oscillator plus an
{\em integer times the parity}. This smacks strongly of the situation
in supersymmetric Hamiltonian models and it would not be surprising
if such an interpretation can be found.

\acknowledgments
I would like to thank the University of Illinois for hospitality
during October 1994, when much of this work was carried out. This
research has been supported by a grant from the Australian Research
Council.


\begin{references}
\bibitem{f1} We could of course have reversed the sign
 of $Q$ or multiplied $D$ by $-iQ$; this would not affect the
 ensuing analysis in any essential way.
\bibitem{f2} that multiplies $\hbar\omega$, if we reinstitute
 constants.
\bibitem{f3} One could of course rotate the sigma matrices to
obtain another equivalent representation.
\end{references}
\end{document}